\documentclass[twoside]{article}
\usepackage[a4paper]{geometry}
\usepackage[latin1]{inputenc} 
\usepackage[T1]{fontenc} 
\usepackage{RR}
\usepackage{hyperref}
\RRNo{2012--04-001}
\RRdate{April 2012}

\RRauthor{
François Durand
\thanks[sfn]{\ALBLF}
  \and
Fabien Mathieu\thanks{\Inria}
\and Ludovic Noirie\thanksref{sfn}
}
\authorhead{Durand \& Mathieu \& Noirie}
\RRtitle{Manipulabilité des systèmes de vote :\\ Applications aux réseaux inter-opérateurs}
\RRetitle{On the Manipulability of Voting Systems: \\ Application to Multi-Carrier Networks}
\titlehead{Manipulability of Voting Systems: Application to Multi-Carrier Networks}
\RRresume{
Aujourd'hui, la plupart des acteurs d'Internet ont des intérêts financiers (fournisseurs d'accès, de service, \ldots). Dans ce contexte, il est crucial de pouvoir prendre des décisions équitables, l'environnement étant très compétitif. En particulier, il s'agit de pouvoir empêcher une partie des acteurs de \emph{truquer} un résultat au détriment des autres, ce qui se fait d'habitude en considérant un point de vue de \emph{théorie des jeux}.

Or, les systèmes de vote présentent une alternative intéressante peu étudiée dans ce contexte, tout du moins à notre connaissance. Ils permettent à des entités concurrentes de choisir en commun entre plusieurs options. Des résultats théoriques classiques montrent hélas que tous les systèmes de vote sont susceptibles d'être manipulés sous certaines conditions, à l'exception de quelques systèmes \emph{dégénérés}, comme par exemple la dictature. Cependant, on ne sait pas grand chose de la manipulabilité quantitative des systèmes de vote pour des scénarios pratiques.

Dans cet article, nous faisons une étude empirique de l'utilisation des systèmes de vote pour choisir une route dans un réseau inter-opérateur, en comparant simultanément l'efficacité du résultat sincère, la manipulabilité et l'efficacité du résultat manipulé. Nous montrons qu'un système en particulier, \emph{Single Transferable Vote} (STV), est bien plus résistant que le système qui donne la solution optimale si tout le monde est sincère, tout en conservant une efficacité proche de l'optimal, que les acteurs essaient de manipuler le système ou non.
}
\RRabstract{
Today, Internet involves many actors who are making revenues on it (operators, companies, service providers,...). It is therefore important to be able to make fair decisions in this large-scale and highly competitive economical ecosystem. One of the main issues is to prevent actors from manipulating the natural outcome of the decision process. For that purpose, game theory is a natural framework.

In that context, voting systems represent an interesting alternative that, to our knowledge, has not yet been considered. They allow competing entities to decide among different options. Strong theoretical results showed that all voting systems are susceptible to be manipulated by one single voter, except for some ``degenerated'' and non-acceptable cases. However, very little is known about how much a voting system is manipulable in practical scenarios. 

In this paper, we investigate empirically the use of voting systems for choosing end-to-end paths in multi-carrier networks, analyzing their manipulability and their economical efficiency.  
We show that one particular system, called \emph{Single Transferable Vote} (STV), is largely more resistant to manipulability than the natural system which tries to get the economical optimum. Moreover, STV manages to select paths close to the economical optimum, whether the participants try to cheat or not.}
\RRmotcle{Systèmes de vote, théorie des jeux, manipulabilité, réseaux inter-opérateurs}
\RRkeyword{Voting systems, game theory, manipulability, multi-carrier networks}
\RRprojets{\ALBLF, \Inria}

\usepackage{amsfonts,amsmath,amstext,amssymb}

\def\R{\mathbb{R}}
\usepackage[caption=false,font=footnotesize]{subfig}
\usepackage{epstopdf}
\usepackage{color}
\usepackage{ifthen}
\newboolean{showcomments}
\setboolean{showcomments}{true}
\ifthenelse{\boolean{showcomments}}
{ \newcommand{\mynote}[3]{
    \fbox{\bfseries\sffamily\scriptsize#1}
    {\small$\blacktriangleright$\textsf{\emph{\color{#3}{#2}}}$\blacktriangleleft$}}}
{ \newcommand{\mynote}[3]{}}

\begin{document}
\makeRR   


\section{Introduction}



The emergence of new services, along with the constant growth of existing ones, has led to the explosion of Internet traffic.
The network of networks has become a huge economical ecosystem, on which lot of companies are making revenues. This includes network operators, that interconnect their infrastructures to make the Internet, but also service providers, that monetize their services on it. In this context, it is important to ensure fairness in decisions that involve many different competing actors, with the goal to reach some kind of global economical optimum. 

Indeed, participants of a decentralized network often have to take global decisions based on local interests. For instance, Internet routes span over multiple Autonomous Systems, while they result from local, arbitrary decisions. Many fields related to game theory have been proposed in order to have a better understanding of distributed decision-related issues (\cite{nisan2007}). 
However, to the best of our knowledge, one of these fields remain mostly unexploited: voting systems, that allow competing entities to decide among different options.


The main goal of this paper is to investigate the usage of voting systems within the Internet economical ecosystem. In particular, we focus on the question of manipulability, which is crucial in a decision-making context: how hard is it for participants to change the decision by lying about their own interests? It is known that except for a few degenerated cases, all voting systems can be cheated (\cite{gibbard73manipulation,satter75strategy,gibbard77manipulation}). However, the practical performance of existing algorithms is mostly limited to high level properties. This study applies the voting system framework to a networking use-case, and quantifies manipulability and its effects on such a scenario. 
This paper is organized as follows.
Section \ref{sec:modeling} gives a generic, self-contained, framework of voting systems, with a focus on manipulability aspects. 
Section \ref{sec:intercarrier-model} defines and models the networking use-case of our paper: path establishment in multi-carrier networks. 
Lastly, in Section \ref{sec:intercarrier-results}, we analyze the results we obtain, illustrating how and why voting systems can be very interesting for practical use in an economical ecosystem such as the one of multi-carrier networks.

\newcommand{\perm}{\mathfrak{S}}
\newcommand{\reel}{\mathbb{R}}

\section{Generalities on voting systems}
\label{sec:modeling}

This section presents the general framework of voting systems that we use in this study. It first describes the notion of elector's preferences, the general definition of voting systems and some examples that we use in our study. Manipulability criteria are then given, as well as some general results that are known about manipulability of voting systems.   

\subsection{Preferences}

We denote $\mathcal{E} = \{e_i\}_{i=1}^{n}$ the set of electors (voters) and $\mathcal{C} = \{c_j\}_{j=1}^{m}$ the set of candidates. 
Following Von Neumann-Morgenstern approach \cite{1947vonneumann}, the preferences of an elector $e_i$ are represented by a utility vector $\mathbf{U}_i = (u_{i,j})_{j=1}^{m}$, with:
\begin{itemize}
\item	$u_{i,j} > u_{i,k}$ means that $e_i$ strictly prefers $c_j$ to $c_k$, 
\item	$u_{i,j} = u_{i,k}$ means that $e_i$ values $c_j$ and $c_k$ in the same way,
\item For $\lambda \in [0,1]$, $u_{i,k} > \lambda u_{i,j} + (1-\lambda) u_{i,l}$ means that $e_i$ strictly prefers $c_k$ to a lottery where $c_j$ is chosen with probability $\lambda$ and $c_l$ with probability $1-\lambda$.
\end{itemize}
Thus, $\mathbf{U}_i$ does not only represent elector $e_i$'s order of preferences but also the compared strengths of her preferences. 

We denote $\mathcal{U} = \R^m$ the space of the possible preference profiles for an elector\footnote{In the vanilla Von Neumann-Morgenstern approach, utilities are defined up to two constants, and $\mathcal{U}$ corresponds to a quotient space of $\R^m$ by the corresponding equivalence relation. In this paper, however, utilities correspond to an actual gain or loss, so we can discard the equivalence relation.}.
Knowing elector $e_i$'s utility vector $\mathbf{U}_i\in\mathcal{U}$, we can deduct $e_i$'s order of preferences: her preferred candidate(s) is (are) the one(s) she attributes the highest utility, etc. This define a canonical surjection $\sigma$ from the utility space $\mathcal{U}$ to the set of weak orders over the candidates.

\subsection{Voting system definition}

A voting system allows competing entities (the electors) to select one option among many (the candidates).
An elector $e_i$ can choose a deterministic \emph{strategy} or \emph{ballot} from a strategy set $S_i$.
This would be called a \emph{pure strategy} in usual game theory.
Once each elector has chosen a strategy, the voting system picks out a candidate, applying a voting rule function:
$$
f : S_1 \times \ldots \times S_n \rightarrow \mathcal{C}.
$$
In order to link utilities to voting systems, we also need to define sincerity: for each $i$, we assume that there is a function
$$
g_i : \mathcal{U} \rightarrow S_i
$$
that describes the ``spirit'' of the voting system: if your preferences are $\mathbf{U}_i$, you are supposed to vote $g_i(\mathbf{U}_i)$ and doing so will be considered \emph{sincere}. Actually, most voting systems admit a simple, canonical, sincerity function:
\begin{itemize}
	\item when $S_i$ is equal to $\mathcal{U}$, we use $g_i = Id$;
	\item when $S_i$ is the set of weak orders over the candidates, we use $g_i = \sigma$, the canonical surjection.
\end{itemize}

Note that a great part of the literature limits \emph{voting systems} to the case where $S_i$ is even the set of strict orders over the candidates~\cite{satter75strategy,1983barbera,2000benoit,2001reny,2009weber}.

\subsection{Examples of voting systems}
\label{subsec:exVotingSyst}

Voting systems are a huge family, and studying all of them is far beyond the scope of this paper. For reasons that will be detailed in Section \ref{sec:intercarrier-model}, we focus on the following systems:

\subsubsection{Range voting} 
each elector $e_i$ communicates a vector of notes, $\mathbf{S}_i = (s_{i,j})_{j=1}^{m}$. The candidate $c_j$ who maximizes the sum $\sum_{i=1}^{n} s_{i,j}$ wins. We consider that $e_i$ is sincere when she communicates her utility vector $\mathbf{U}_i$ as vector of notes. In order to limit obvious ways of cheating, we can give restrictions to the vector $\mathbf{S}_i$: for example, a minimum and a maximum value allowed for each coordinate.

\subsubsection{Exhaustive ballot (EB)} the protocol proceeds through a series of $m-1$ elimination rounds. At the beginning of each round, each elector communicates her preferred candidate among the remaining ones\footnote{In the generic descriptions of Exhaustive ballot and STV, we assume that for each elector, the order of preferences is strict. If an elector can be indifferent between several candidates, these system need further precisions, as the ones we propose in section \ref{voting-systems-network}}, and the candidate with least votes is eliminated. The winner is the last remaining candidate.

\subsubsection{Single Transferable Vote (STV)} each elector communicates her order of preferences once and for all. Then, the protocol emulates a series of $m-1$ elimination rounds. In each round, the candidate who is ranked first by the least number of voters is eliminated and each ballot in her favor is transferred \emph{automatically} to the best ranked remaining candidate in that ballot. 

It is straightforward that STV is indeed an emulation of EB. In particular, if all electors vote sincerely, both systems give the same result. The main difference is that in EB, an elector can contradict herself from one round to another, like changing the candidate she says she prefers between two rounds even if she has not been eliminated\footnote{In that case, she uses a strategy that is not in the image of her sincerity function $g_i$. It is obviously not sincere, but allowed in this voting protocol.}.

\subsection{Manipulability criteria}

We need properties that describe the manipulability of voting systems.
We consider a given set of electors, candidates, utilities, and a voting system with an associated sincerity function. $v \in \mathcal{C}$ denotes the candidate who is elected when all electors vote sincerely. The following definitions hold.



\subsubsection{Coalition manipulability (CM)} 
a subset of voters, by casting insincere ballots, can make the result of the vote better (from their point of view). That is, there exists a challenging candidate $c \in \mathcal{C}\setminus \{v\}$ such that the electors preferring $c$ to $v$ can cast their ballots so that $c$ gets elected, assuming that other electors don't change their own ballots.

\subsubsection{Trivial coalition manipulability (TM)} there is a challenging candidate $c\in \mathcal{C}\setminus \{v\}$ that gets elected if all electors preferring $c$ to $v$ use their \emph{trivial strategy}. Intuitively, an elector $e$ uses her trivial strategy for candidate $c$ against candidate $v$ when she pretends that all candidates in $\mathcal{C} \setminus \{v, c\}$ keep their sincere utility, 
that $v$ has a finite utility lower than theirs and that $c$ has an infinite utility.
For instance, the trivial strategy in range voting consists of attributing the maximum note to $c$ and the minimum note to other candidates. In STV or Exhaustive ballot, the trivial strategy means putting $c$ at the top and $v$ at the bottom, while keeping one's sincere, personal order of preferences over the other candidates.




CM is about being able to manipulate the system, not matter how hard it is. It answers the question: can the system be manipulated by omnipotent manipulators, which have a complete knowledge of the system, and can cast coordinated ballots? TM, on the other hand, answers the question of a zero-knowledge, decentralized, manipulation: once the challenger $c$ is chosen, all electors of the coalition can cast their own ballot independently.


\subsection{Gibbard-Satterthwaite theorem}

We say that an elector $e_i$ is a \emph{dictator} iff, for any possible outcome $c \in f(S_1\times\ldots S_n)$, elector $e_i$ can cast a ballot so that the elected candidate will be $c$, whatever other electors vote. We say that a voting system is \emph{dictatorial} if there is a dictator.

It is \emph{manipulable} iff at least one single elector can benefit from casting an insincere ballot (which implies coalition manipulability (CM)).

The following theorem was proved by Satterthwaite \cite{satter75strategy} for voting systems based on permutations and by Gibbard \cite{gibbard73manipulation} in the general framework of strategies:

\emph{Every non-manipulable voting system with at least three possible outcomes is dictatorial.}

This result can indeed be generalized beyond the assumptions that we made here,
like 
for instance if the voting system is non-deterministic \cite{gibbard77manipulation}.


\section{Multi-carrier networking case}
\label{sec:intercarrier-model}

In this section, we define the multi-carrier end-to-end path establishment problem we want to solve. We first explain how voting systems can be applied for this purpose, and how the set of candidate paths is selected. Then we define the cost and gain modeling, how the voting systems are applied and how the manipulability is calculated.  

\subsection{Voting systems for multi-carrier path establishment}

The problem to solve is the following: in a multi-carrier network, for given ingress and egress, which end-to-path should be selected if one takes carriers' preferences into account.


\begin{figure}[!t]
\centering
\includegraphics[width=0.6\textwidth]{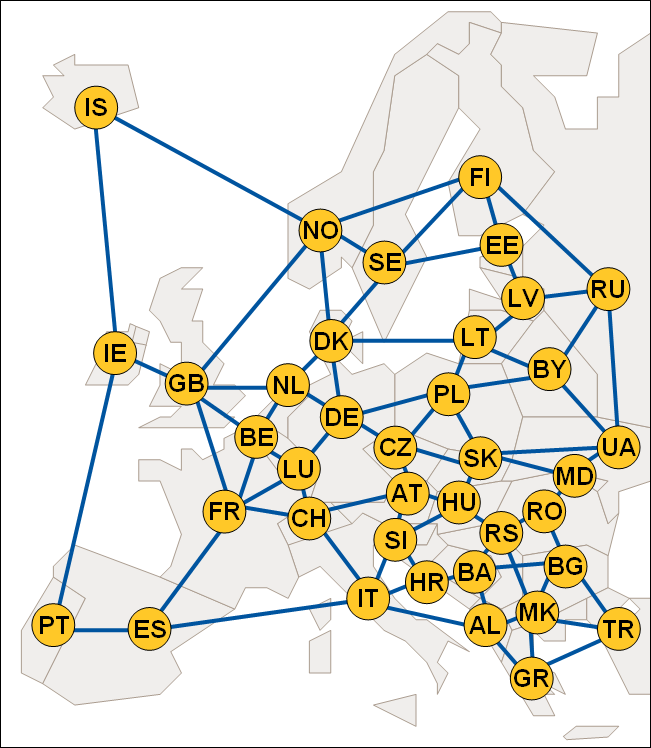}
\caption{Multi-carrier network example}
\label{fig_icn}
\end{figure}

Figure \ref{fig_icn} represents an example of interconnection of multi-carrier networks, with one carrier network per European country, interconnected by some geographical neighborhood. 

We are aware that this example is not representative of a real multi-carrier network. For instance, such a flat interconnection topology differs from the historical BGP hierarchical topology of Internet\footnote{Note that some recent studies show an evolution towards flatter topologies (\cite{Labovitz2010,Dhamdhere2010}).}. Nevertheless, our goal here is to give a simple model where costs derive from some sort of underlying metric, so a geographical basis is a natural choice.


Because the carriers are in competition and can express their preferences, voting systems are well-adapted to model the decision process in the end-to-end path selection. We propose the following identification between the inter-carrier problem and the voting system modeling: 
\begin{itemize}
	\item The electors/voters $\mathcal{E}$ are the carriers. In our present study, we only consider one elector per carrier, but one may envisage some variants (e.g., 
	give different weights to different carriers, include the end-users, etc).  
	\item The candidates are the feasible routing paths for a given demand (or a subset of them).	They are an input of the problem: issues related to	their identification or Quality of Service (QoS) are out of the scope of this study.
	\item The preferences of each carrier are represented by a utility vector, 
	from which a ballot is defined (cf \S \ref{sec:modeling}).   
	\item The election result is the selected path for the client request. Any voting system can be used. Here we suppose there is an independent entity, called \emph{supervisor}, in charge of the election process. The supervisor collects the ballots from all the carriers and processes the voting algorithm to decide the winner path (other options are possible: e.g., the carriers cooperatively participating to the whole voting scheme). Ensuring the integrity of the supervisor is beyond the scope of this paper.
\end{itemize}

Some carriers may have some knowledge about the utilities of their competitors:
\begin{itemize}
\item by public knowledge about the utilities;
\item by cooperation between some carriers (within a coalition); 
\item by inference on previous votes (learning);
\item by spying or information interception. 
\end{itemize}
So the carriers may use that knowledge to lie about their own preferences, 
 in order to improve their benefits, maybe against the global interest. Studying the manipulability of the voting systems is thus important in that context. 

\subsection{Candidate paths}
In our study case, we are considering the interconnection of $n=38$ carrier networks of figure \ref{fig_icn}.

A \emph{demand} is a request for end-to-end connections, with the only constraint that the connection must start at a given carrier network (ingress) and end at another one (egress).

For each demand, we could consider the set of possible paths without loop. However, the size of that set increases exponentially with the number of nodes in the network, so we want the supervisor to limit the number of candidates.

The only information the supervisor should use for selecting candidates is the multi-carrier interconnection topology, because carriers may wish to avoid any cost information diffusion before the actual vote, the rule we applied to limit the number of candidate paths is the following:
\begin{itemize}
\item The supervisor fixes a minimal number of candidate paths $m_{min}$, and an initial threshold $\delta_{min}$.
\item It computes the minimal number of hops $h_{min}$ to satisfy the demand.
\item It takes all paths without loop with a number of hops less than or equal to $h_{min}+\delta_h$, where $\delta_h$ is the smallest threshold that is greater than or equal to $\delta_{min}$ and that selects at least $m_{min}$ candidates.
%
\end{itemize}  

It is important to note that:
\begin{itemize}
\item The candidate paths are fully determined by the demand and the parameters $(m_{min},\delta_{min})$. Their exact number depends on the demand.
\item For each candidate path $c_j$, only a subset $\mathcal{E}_j$ of the carriers $e_i \in \mathcal{E}$ are concerned by the candidate path.
\item Among these carriers, some may be concerned by only a subset of candidates.
\end{itemize} 

For numerical evaluations, we use two limitation options: $(m_{min},\delta_{min}) \in \left\{ (5,0); (10,1) \right\}$. The first option gives on average 9.94 candidate paths per request (from 5 to 43), and the second option gives on average 21.25 paths per request (from 10 to 127).

\subsection{Multi-carrier cost and gain modeling}
\label{subsec:gainmodeling}

We need to define the utilities in order to derive the preferences and ballots of the carriers.
From an economical point of view, the natural definition is to take the difference between the gains and costs that carriers have for each possible paths. 

As the focus of our study is the interest of voting systems, we limit ourselves to a very simplified cost and gain model.

\subsubsection{Cost}

The cost for a carrier $e_i$ in $\mathcal{E}_j$ to carry the path $c_j$ is noted $\alpha_{i,j}$. There are numerous ways to value $\alpha_{i,j}$, but for the sake of simplicity, we define $\alpha_{i,j}$ as the sum of half the cost of the incoming interconnection link (which is null for the ingress carrier) and half the cost of the outgoing interconnection link (which is null for the egress carrier). For the cost of the interconnection link between two adjacent carriers $a$ and $b$, we choose a linear function $C_0 + d_{a,b} / d_0$ of the distance $d_{a,b}$ between $a$ and $b$ (we used the distances between the capital cities in our multi-network example). In our numerical study, we considered three cost options: dominated by the constant cost $C_0$ ($C_0 = 1$ and $d_0 = 100\times\max(d_{a,b})$), purely linear ($C_0 = 0$, $d_0 = \max(d_{a,b})$), and intermediate ($C_0 = 1$, $d_0=\max(d_{a,b}) / 3$). 


\subsubsection{Gain}

Concerning the gain for the carriers, we consider that the client will pay a fixed amount $A$ for a given demand
 (flat cost model for the client point of view). If path $c_j$ is selected, this amount $A$ will be equally distributed between the concerned carriers (the ones in $\mathcal{E}_j$): 
\begin{itemize}
\item This flat-fare approach is similar to what happens for some city metropolitan transportation systems, where the ticket price is a fixed fare not depending on the distance between the departure and the arrival stations. Other choices are of course possible, but the important assumption is that 
the price to pay for a connection is known in advance. In our study, with a flat rate, the benefit for the operator to set up the connection may be negative for long paths, but on average, concerning short and long paths, he will obtain a positive benefit. 
\item The client fare may be non-equally distributed. But the question on how the revenue is distributed between the carriers is out of the scope of our study. For the sake of simplicity, we thus choose the equal repartition.  
\end{itemize}

We fixed the value $A$ of the user fare in such a way that, considering the least cost paths for each request, the average global revenue represents 140\% of the average global cost (i.e., benefit of 40\%). How to decide $A$ in the real world is out of the scope of this study that focuses on manipulability.

\subsubsection{Utility}
The sincere utility value for carrier $e_i$ to carry the candidate path $c_j$ can be defined as the net income (positive or negative) for the carrier if this candidate path is selected:
\begin{itemize}
\item	$u_{i,j} = A / card(\mathcal{E}_j) - \alpha_{i,j}$ if $e_i \in \mathcal{E}_j$ 
\item	$u_{i,j} = 0$ if $e_i \notin \mathcal{E}_j$ (the carrier is then \emph{indifferent})
\end{itemize}


\subsubsection{Global income}

The global income of a given carrier depends on the intensity and distribution of demands. While intensity only scales the result, the distribution can change its nature.
In our study, we consider that demands are uniformly distributed among each pair of different ingress and egress carriers. Of course, results can be straightforwardly generalized to any kind of distributions, related to country population or to some traffic matrix.

\subsection{Voting systems for the multi-carrier study case}\label{voting-systems-network}

\subsubsection{Range voting} With the utilities previously defined, the most natural choice for a voting system is the one that maximizes the global revenue, which is the sum of the utilities of all carriers contributing to the selected path. This corresponds exactly to the \emph{range voting} defined in Subsection \ref{subsec:exVotingSyst}, so \emph{range voting} is the natural reference system. In details, with range voting:
\begin{itemize}
\item	The voters (carriers) give their utilities to the supervisor.
\item	The supervisor sums the notes for each candidate path, and selects the path with the maximal value.
\end{itemize}
One of the drawbacks with \emph{range voting} is that the carriers are required to give all the information about their cost to the election supervisor. Even if it is an independent entity, they may wish to avoid giving this kind of information.

\subsubsection{STV} Numerous other voting systems may be applied. 
Previous work, as \cite{chamberlin1984} or \cite{walsh2010empirical}, suggests that STV belongs to the least manipulable voting systems known, so we used STV as the second voting system for our use case\footnote{Actually, we also tried other voting systems in experiments not presented here, and verified that STV was the most promising candidate with respect to manipulability.}. Path selection with \emph{STV} works as follows:

\begin{itemize}
\item	The voters (carriers) give the order of preferences on the candidate paths with which they are concerned. But the supervisor must also know how to position the other paths with which they are not concerned and to which they must be indifferent, so the carriers must also say to the supervisor, for each candidate path with which they are concerned, if they like it (financial gains) or dislike it (financial losses). They can deduce all the required information from their utilities, but they do not need to transmit the actual value of their utilities, which may be preferable for the carriers.
\item	The supervisor processes the \emph{STV} mechanisms as defined in Subsection \ref{subsec:exVotingSyst}, with the following rule with equal preferences: in each round of \emph{STV}, for a given carrier (elector), in case the number of remaining candidate paths with the highest preference is equal to or higher than two, then the vote is equally divided between these candidates. In each round, the election supervisor eliminates the candidate path with the least votes (the number of votes may be not an integer value in this case). In case of ties, we choose to eliminate the candidate path with the lowest index (it could be done randomly, but the impact is marginal in our problem). 
\end{itemize}



\subsection{Manipulability algorithms}
\label{subsec:manipAlgo}

First, for both voting systems, the manipulation of the vote by a carrier is limited to the candidate paths concerning this carrier: the carrier cannot pretend to like or dislike a candidate path to which he must be indifferent.

\subsubsection{Range voting} In range voting, when electors preferring $c$ to $v$ try to make $c$ win, their best strategy is obviously the trivial one: give the maximum note to $c$ and the minimum note to the other candidates.
For our use case, the only adaptation we made is that a carrier must give a null utility to the paths it does not belong to (this can be enforced by the supervisor, which knows the topology), but it is still the best strategy to make $c$ win. Because of that, TM and CM are equivalent. The maximum note is set to $+A$ and the minimum note to $-A$ in our study.

\subsubsection{STV}
Finding out if there is a way to manipulate with STV is much more difficult: in fact, the problem is known to be NP-hard \cite{bartholdi1991}. To limit this difficulty, we use simple methods that test the manipulability (or non-manipulability) of a given demand. When the tests are inconclusive for a given demand, we can only answer \emph{maybe}, but when considering all demands, this allows to give lower and upper bounds for manipulability.

In order to prove that STV is manipulable, we just try trivial manipulation. That is, for any candidate $c$, electors prefering $c$ to $v$ pretend that $c$ is their best choice, $v$ is their worst choice, keeping other candidates in their sincere positions. This gives a lower bound of the manipulability.

In order to prove that STV is not manipulable, our algorithm is an adaptation of \cite{coleman2007}. The idea is to use a variation of the system that gives more power to manipulators and for which CM can be exactly computed. If the altered system cannot be manipulated, STV cannot either.
In details:
\begin{itemize}
	\item At each round, we authorize the manipulators to change their vote. This shifts the voting paradigm from STV to Exhaustive ballot. The interest is that eliminating candidate $a$ then $b$ or $b$ then $a$ lead to the same situation, whatever the manipulators have done to get there. This permits an iterative approach instead of a recursive one.
	\item At each round, each manipulator can share her vote between several candidates, even non equally, for instance $\frac{1}{3}$ vote for one candidate and $\frac{2}{3}$ vote for another one. This allows a water-filling approach and avoid Knapsack-like issues.
	\item We authorize electors to lie even about the paths they don't belong to. So all manipulators are symmetric in right and we don't have to manage them individually.
\end{itemize}
With these modifications, we can manage the group of manipulators globally: at each round, we can divide their votes as we like between the candidates, in order to eliminate the candidate we want to.

When we prove that manipulation is impossible with these adapted rules, then it is impossible with the rules given in \ref{voting-systems-network}: this provides an upper bound for the manipulability of STV.

As we will see in next Section, lower and upper bounds tend to be close to each other, so we get a good estimate of manipulability.


\paragraph*{Remark} Actually, the previous algorithm is still very costly (virtually in $n 2^m$, where $n$ is the number of voters and $m$ is the number of candidates). When there are more than 25 candidates and that trivial manipulation does not work, we don't try to prove the impossibility and we directly consider the test inconclusive.

\section{Multi-carrier networking results}
\label{sec:intercarrier-results}

In this section, we analyze the results we obtain on the multi-carrier network case described in Section \ref{sec:intercarrier-model}.

For both \emph{range voting} and \emph{STV}, we first zoom on one specific scenario that will serve as reference, and then we extend the results for several parameters. We observe manipulability and economical efficiency, with sincere and insincere preferences.

\subsection{Reference scenario}

We choose for reference scenario the case with $(m_{min},\delta_{min})=(5,0)$ (on average $9.94$ candidate paths per demand) and the intermediate link cost model ($C_0 = 1$, $d_0=\max(d_{a,b}) / 3$). For this scenario, we measure:
\begin{itemize}
	\item \emph{Sincere efficiency}: global net income, in percentage of the optimal global net income, assuming that all carriers are sincere.
	\item \emph{Manipulability}: proportion of the demands that are CM. For range voting, this can be exactly computed because CM is TM. For STV, we use TM as a lower bound and the Exhaustive Ballot variant to give an upper bound (cf \S \ref{subsec:manipAlgo}).
	\item \emph{Insincere efficiency}: efficiency of the system when manipulations are allowed. For a given demand, several manipulations can occur. We can measure the average insincere efficiency, if one manipulation is selected at random, or the worst case situation if for each manipulable demand, one chooses the manipulation that minimize the global net income (remember that by definition, the net income of the manipulators will increase, though).
\end{itemize}

\begin{table}%
\begin{center}
\begin{tabular}{|p{5cm}|c|c|}
\hline
Voting system & Range Voting & STV \\
\hline
Sincere efficiency & 100\% & 95 \% \\
\hline
Manipulability & 96\% & $<$ 20\% (TM: 18\%)\\
\hline
Insincere efficiency (average) & 37\% & 90\% \\
\hline
Insincere efficiency (worst case) & -75\% & 89\% \\
\hline
\end{tabular}
\caption{Main results for the reference scenario ($(m_{min},\delta_{min})=(5,0), (C_0,d_0)=(1,\max(d_{a,b}) / 3)$ )}
\label{tab:reference}
\end{center}
\end{table}


The results are presented in Table~\ref{tab:reference}. They clearly indicate that STV present many advantages over the more natural range voting selection:
\begin{itemize}
	\item For sincere voting, i.e. with no manipulation, \emph{range voting} manipulability gives the economical optimum as expected. While not having global optimum as a target STV still manages to achieve about 95\% to the optimum. This slight efficiency decrease can be seen as the price to pay for robustness (see below).
\item Almost all demands (96\%) can be manipulated if ranged voting is used, against less than 20\% for STV. 
\item Considering the impact on efficiency, the degradation is very high for \emph{range voting}: it becomes only 37\% of the optimum on average, down to -75\% in the worst case scenario. On the other hand, STV maintains a robust $90\%$ on average ($89\%$ in the worst case scenario).
\end{itemize}

So we should retain that STV is much less subject to manipulation than range voting, and that even when a manipulation exists, its impact on the global welfare is bearable.



\subsection{Impact of parameters}

We proposed in Section \ref{sec:intercarrier-model} two path-limitation parameters and three cost models, so we have six possible configurations. Results for these six configurations follow.

\subsubsection{Manipulability}

\begin{figure}[!t]
\centering
\includegraphics[width=0.7\textwidth]{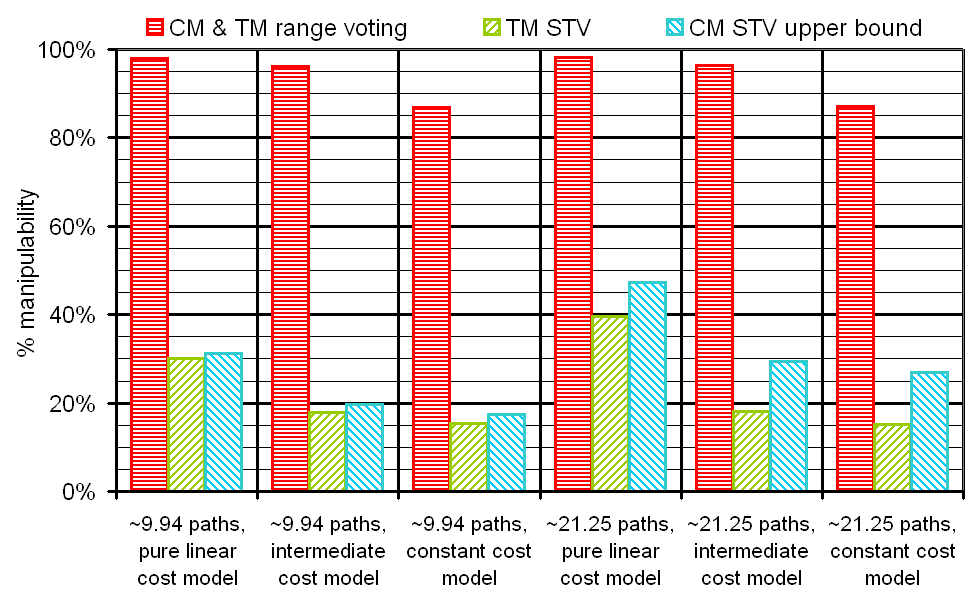}
\caption{Manipulability of STV vs. Range Voting}
\label{fig_manip}
\end{figure}

Figure \ref{fig_manip} indicates the manipulability of the considered scenarios.

For both voting systems, the more numerous the path candidates are, the more manipulable they are. This is somehow expected, as more candidates mean more possible challengers for manipulation. 
So the supervisor should limit the number of proposed candidates to decrease the manipulability, while keeping enough candidates to allow a fair path selection. Concerning the link cost model, we observe that the flatter it is, the lower the manipulability (for both voting systems). But the most interesting result is that the manipulability of STV stays much lower than the one of \emph{range voting} in all cases. 
While \emph{range voting} manipulation is always higher than 85\%, STV manipulation is about 30\% or less except for one scenario (between 40\% and 48\% for the highest number of candidate paths with purely linear link cost model).

Note that for STV, the lower and upper bounds for manipulability are relatively close. The differences are higher when the number of candidate paths increases, but this is due to the way we calculate the upper-bound (see Section \ref{sec:intercarrier-model}), skipping the evaluation for demands with too many candidate paths. 


\subsubsection{Sincere economical efficiency}  



\begin{figure}[!t]
\centering
\includegraphics[width=0.7\textwidth]{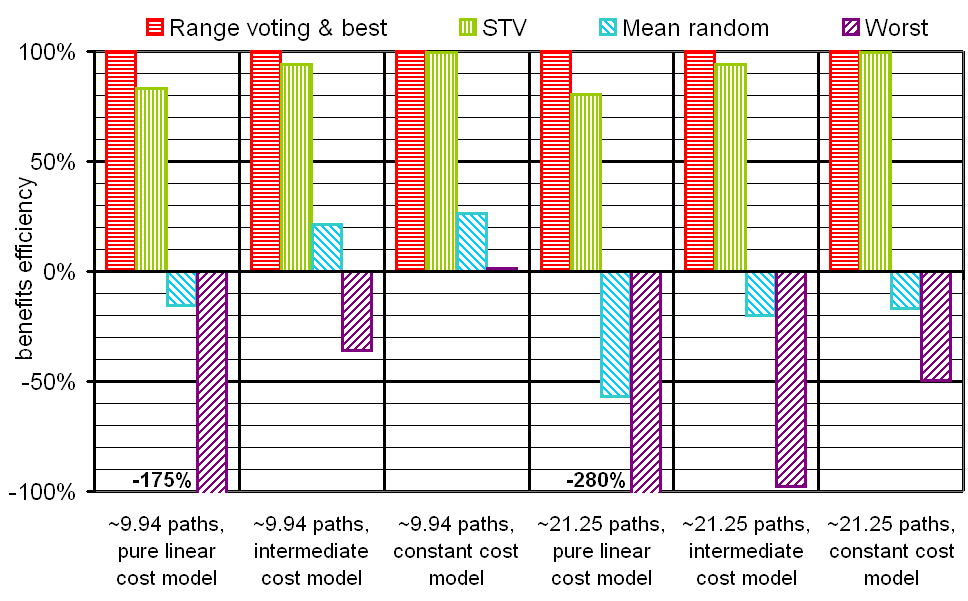}
\caption{Efficiency of STV vs. Range Voting for sincere preferences}
\label{fig_eff_sincere}
\end{figure}

Figure \ref{fig_eff_sincere} gives the sincere efficiency of the manipulability of the considered scenarios.
As expected, \emph{range voting} gives 100\% economical efficiency. But STV gives an economical efficiency close to this optimum. For both low or high number of candidate paths, this efficiency is about 80\% for purely linear link cost model, about 95\% for intermediate link cost model, and more than 99\% for constant link cost model. This confirms that, even if STV may give a path which is not the optimal one for global economical benefits, the selected path is quite close to the optimal choice. 

For completeness, Figure \ref{fig_eff_sincere} also indicates the efficiency obtained when the path is chosen at random among the candidates, and when the worst candidate is chosen.


\subsubsection{Insincere efficiency}



\begin{figure}[!t]
\centering
\includegraphics[width=0.7\textwidth]{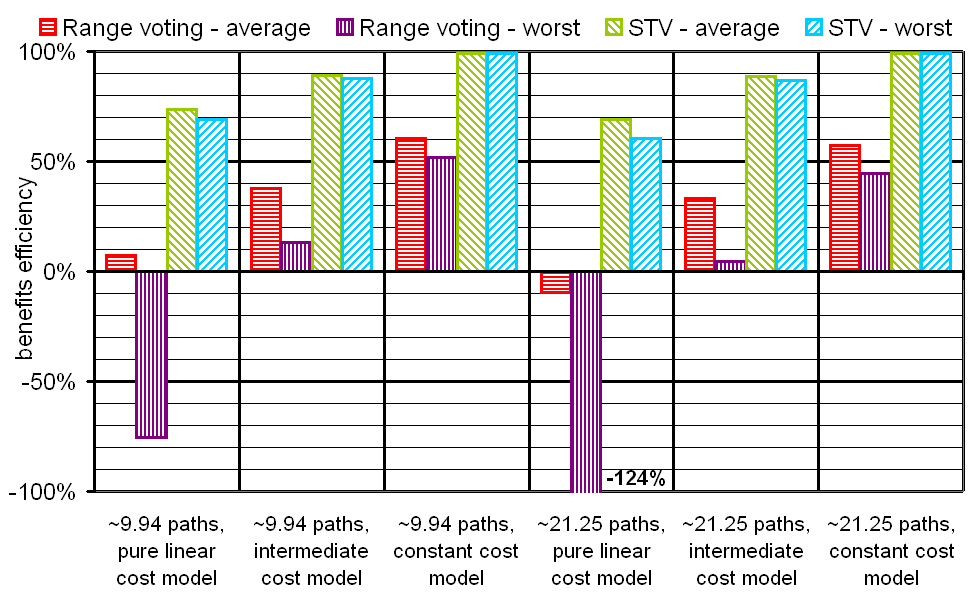}
\caption{Efficiency of STV vs. Range Voting for manipulated preferences}
\label{fig_eff_manip}
\end{figure}

To complete our study, Figure \ref{fig_eff_manip} displays the insincere efficiencies.
For all scenarios, one observes a huge gain by choosing STV instead of \emph{range voting}. For STV, the economical efficiency is only slightly degraded compared to sincere preferences, and the worst case is never far from the average case. Numerically, insincere efficiencies range between 60 to 70\% for the purely linear link cost model (vs. about 80\% for the sincere preferences), about 90\% for the intermediate link cost model (vs. about 95\% for the sincere preferences), and it is even slightly improved for the constant link cost model (not visible on the graph, above 99\% in all cases).
For \emph{range voting}, the economical efficiency is largely degraded compared to sincere utilities: while the economical efficiency with sincere utilities is 100\%, when considering manipulations, it falls to: about 0\% on average and large negative values on worst case for the purely linear link cost model; about 35\% on average and about 8\% on worst case for the intermediate link cost model; about 60\% on average and about 50\% on worst case for the constant link cost model.

All these results confirm that STV is really safer to preserve the economical benefits of the selected candidate path than \emph{range voting}.

\section{Conclusion}
\label{sec:conclusion}


Existing theoretical results show that, except for a few degenerated cases, 
any other voting system is susceptible, in some scenarios, to be manipulated, even by a single voter.
In this paper, we proposed to quantify manipulability in practical scenarios and to measure its effects with respect to global welfare. We focused on the use case of multi-carrier end-to-end path establishment. We compared two voting systems:
\begin{itemize}
\item \emph{Range Voting}: it corresponds to what a classical game theory approach where participants are supposed to try and optimize the global welfare. Indeed, range voting maximizes the global net income of the system, provided that the carriers (the voters) give their sincere utilities on the proposed candidate paths (their own net income). 
\item \emph{STV}: this is another voting system, based on elimination rounds. STV has a ``reputation'' to be less manipulable, which we were able to validate in our study.
\end{itemize}



In the end, our study highlights the interest of voting systems in the context of Internet economical ecosystems where many competing players are involved, with end-to-end path selection as a practical use case.

We also observed that all voting systems are not equivalent: when carriers cannot be trusted, STV can largely outperform (in our framework) the economical gain maximization that \emph{Range Voting} is supposed to achieve: 
\begin{itemize}
\item STV manipulability probability can be as low as 20\% while, in the same configuration, \emph{Range Voting} is close to 100\%.  
\item With \emph{STV}, the carriers do not need to give all the information to process the vote, the preference order with the indifference limit is enough, while for \emph{Range Voting}, they must give the whole information about their cost. So not only is STV less manipulable, but it also limits the diffusion of knowledge that could be used for manipulation.
\item Although STV does not target the economical optimal choice, it remains very close to that optimum. The price of non-manipulability is low (in the range of ~5\% in our study).
\item Considering manipulations, the degradation on the economical efficiency is limited with STV, while it is huge with \emph{range voting}, which reinforces the interest of STV.
\end{itemize}

\subsection*{Future work}

The work presented in this paper opens many opportunities for future works:  
\begin{itemize}
\item Analysis of the sensibility of various parameters: in our use-case, we started to analyze the sensibility to the link cost structure and number of proposed candidate paths. But lot of other parameters could be tested: changes in the revenue model for the carriers (e.g., user's fare proportional to the length of the path instead of flat fare in our study), other multi-carrier topologies (e.g., more hierarchical topologies as one can have in the today's Internet), etc. 
\item Analysis of other voting systems one can find in the literature, or proposition of new efficient voting algorithms to apply. 
\item Identification of other use cases in the context of Internet, with other kind of ecosystems. Here is a non-exhaustive list of such use cases: service composition at the application layer with many service providers, full service offer involving all the actors of the economical chain from end-users to operators and service providers, selection of master nodes in peer-to-peer networks, etc. 
\end{itemize}
Indeed, the framework of voting systems can be useful in any situations in which one can identify voters that need to decide among different options. We proved in our study case that, by correctly choosing the voting system, one can limit the manipulability by some coalition of voters and preserve the revenue for the global economical ecosystem.

\nocite{*}
\bibliographystyle{plain}
\bibliography{voting}

%
%

\end{document}